\documentclass[12pt]{iopart}

\usepackage{graphicx}

\begin{document}

\date{}
\title {Geometric model of dark energy}
\author{
V.~Folomeev${}^{a}$, V.~Gurovich${}^{b}$ and H. Kleinert${}^{c}$}

\address{${}^{a}$ Physics Institute of NAN KR, Chui str. 265 a, Bishkek,
720071,  Kyrgyz Republic}
\address{${}^{b}$ Physics Department, Technion, Haifa 32000, Israel}
\address{${}^{c}$ Inst. f. Theor. Physik, Freie Univ., Arnimallee 14, D-14195 Berlin, Germany}


\begin{abstract}
A cosmological model with
a gravitational
Lagrangian $L_g(R)\propto R+A R^n$ is
 set up to account for
the presently observed re-acceleration of the universe.
The evolution
equation for the scale factor $a$ of the universe  is analyzed in
detail for the two parameters $n=2$ and $n=4/3$, which were
preferred by previous studies of the early universe. The initial
conditions are specified at a red-shift parameter $z\approx 0$. The
fit to the observable data fixes the free parameter $A$. The analysis shows
that the model with $n=2$ agrees better with present data. Then,
if we set $w(q)=-1$ at $z=0$,
 corresponding to the
deceleration parameter $q\approx -1/2$, we find that at $z\approx
0.5$, $w(q)$ has evolved to $w\approx -0.72$, corresponding to
$q\approx 0$. At $z\approx 1$ we find  $w\approx 0$ corresponding
to $q\approx 1/2$. These results are compared with the flat Friedmann
model with cold matter and Lambda-term (LCDM model) for the same initial conditions at $z\approx 0$.
The other
choice of the model with $n=4/3$ allows for  big crunch. However
this possibility is eliminated by the fit of $A$ to the present
data.

\end{abstract}

\bigskip

\section{Introduction}
The discovery of a re-acceleration in the present expansion of the
universe is interpreted as evidence for the existence of a
cosmological $\Lambda$-term in the present equation of motion. At
present, the most popular explanation is based on quintessence
models which extend Einstein's action by scalar fields with
different potentials. The additional parameters are chosen to
comply with observable data. For a review see~\cite{Sahni,Kamen}.

When studying such models it is important to match them with
presently popular models of the early universe. Also here scalar
fields have been used to modify Einstein's equations. But another
important set of models makes use of curvature terms generated by
fluctuating quantum fields as a Casimir effect, a mechanism
discovered by Sakharov \cite{SAKH}. This gives rise to additional
terms $\Delta L(R)$ which behave quite differently from the
Einstein-Hilbert Lagrangian \cite{ZH,CS}.

Starting from a scale invariant renormalizable action proportional
to $R^2$ it is possible to generate for small $R$ an Einstein
action \cite{SG} plus terms of the form
\begin{equation}
\label{addterm} \Delta L(R)=A R^2 + C R^2 \ln|R^2 /R_{*}^2|
\end{equation}
where  $A$, $C$ are parameters. The effect of such terms is
 studied with general methods developed
an arbitrary function $ \Delta L(R)$ in the Friedmann
models~\cite{Breiz}.

Such additional terms can, of course, be generated alternatively
by an Einstein action coupled to a scalar field $\sigma $ with a
suitably chosen potential~\cite{Barrow, Gur}. For instance, $
\Delta L(R)=AR^2$ is obtained from $ \Delta L(R,\sigma)=\sigma
R-\sigma^2/4A$ by extremization in $\sigma$.

There have been ample discussion of models in which $ \Delta L(R)$
is chosen to fit observable data~\cite{Capoz}. Our choice
(\ref{addterm}) has the advantage having a physical origin and
possessing only three parameters which can be fixed completely by
present data at $z=0$, thus allowing
us to calculate uniquely the
future and past evolution of the universe.

\section{Basic Equations}
\label{basic}
As suggested by observations, we consider the flat cosmological Friedmann model with the metric
\begin{equation}
ds^2=d\tau^2-a(\tau)^2 (dx^2+dy^2+dz^2).
\end{equation}
If $H_0$ denotes the presently observable Hubble constant
 (the
subscript 0 will always indicate the present epoch), the reduced
curvature tensor $\rho_{k}^{l}\equiv H_0^{-2}R_{k}^{l}$ has the
following matrix elements as a function of the reduced time
$\theta\equiv H_0 \tau$, with the notation $ \dot{a}\equiv
da/d\theta$:
\begin{eqnarray}
\label{eqmot}
\rho_{0}^0 &=&-3 \ddot {a}/a,\\
\label{eqmot1}
 \rho_{i}^{i}&\equiv&\rho=-6\left( \ddot
{a}/a+\dot{a}^2/a^2\right).
\end{eqnarray}
The variation of Einstein's Lagrangian with an additional term $
\Delta L(R)\equiv f(R)$ gives
\begin{equation}
\label{einst}
G_{i}^k=\frac{8\pi G}{H_{0}^2}T_{i}^{k}+\hat{T}_{i}^k;
\quad G_{i}^k=\rho_{i}^k-\frac{1}{2}\delta_{i}^k \rho.
\end{equation}
Here $T_{i}^k$ corresponds only to cold matter in the present
universe and
\begin{equation}
\label{T}
\hat{T}_{i}^k=-\left\lbrace \left( \frac{\partial f}{\partial \rho}\right) \rho_{i}^k -
\frac{1}{2}\delta_{i}^{k} f+\left( \delta_{i}^{k} g^{lm}-\delta_{i}^{l} g^{km}\right)
\left( \frac{\partial f}{\partial \rho}\right)_{;l;m}\right\rbrace
\end{equation}
specifies the effective quintessence with the nontrivial
dependence on curvature.

We proceed as in Ref.~\cite{Gur1} by introducing the new variable
\begin{equation}
\label{ya} y\equiv(\dot{a} a)^2
\end{equation}
which allows to reduce the order of the equations. Then the
$i=k=0$ component of Eq. (\ref{einst}) leads to
\begin{equation}
\label{00} y+\left[ f_{,\rho}\left(
y-\frac{a}{2}\frac{dy}{da}\right) -\frac{a^4}{6}f(\rho)+a y
\frac{df_{,\rho}}{da} \right] =\frac{\rho_c}{\rho_{*}}
a^4\equiv\Omega_c a,
\end{equation}
where $\rho_c$ is the $a$-dependent cold dark matter (CDM) energy
density, $\rho_*=3 H_0^2/8\pi G$ is the critical density. Choosing
the value of the scale factor $a(\theta_0)$ equal to 1 at present,
one has $\rho_c=\rho_0/a^3$ and
$(\rho_c/\rho_{*})a^4\equiv\Omega_c a$.

In order to investigate the evolution of the cosmological model it
is enough to obtain the solution of Eq. (\ref{00}) with
appropriate initial conditions. But for interpretation of the
solution and for its comparison with observations it is necessary
track for changes of CDM energy density
 $\rho_c$ and quintessence energy density $\rho_v$
separately. For this purpose we use $i=k=0$ component of
Einstein's equations (\ref{einst})
\begin{equation}
G_0^0=\rho_0^0-\frac{1}{2}\rho=\frac{8\pi G}{H_0^2}\left( \rho_c+\rho_v \right).
\end{equation}
Multiplying this equation by $a^4$ and using Eqs.~(\ref{eqmot})
and (\ref{ya}), we have
\begin{equation}
y=(\rho_c+\rho_v) a^4/\rho_{*}
\end{equation}
accounting for the evolution of cold matter energy density. From
this we find for the quintessence energy density:
\begin{equation}
\frac{\rho_v}{\rho_{*}}=(y-\Omega_c a)/a^4.
\end{equation}
By solving Eq.~(\ref{00}), we
 can find the evolution of $\rho_v$. At
known $y$, the Hubble parameter and the curvature components are
obtained from
\begin{equation}
h(a)=\sqrt{y/a^4}; \quad \rho=-\frac{3}{a^3}\frac{dy}{da}; \quad
\rho_{0}^0=3\left( y-\frac{a}{2}\frac{dy}{da} \right).
\end{equation}
We now use  Eqs.~(\ref{eqmot}) and~(\ref{eqmot1}), rewritten as
\begin{equation} \frac{\rho}{6}+H^2=-\frac{\ddot a}{a},
\end{equation}
to derive an expression for the deceleration parameter of the
universe
\begin{equation}
q\equiv -\frac{\ddot a a}{\dot a^2}
\end{equation}
in terms of the curvature components as follows:
\begin{equation}
q=\frac{a^2}{\dot a^2}\frac{\rho}{6}+1.
\end{equation}

For a comparison with observable data we relate
 the parameter $q$
to quintessence and CDM energy densities. Let us suppose that the
quintessence pressure and energy density are proportional to each:
$p_v=w \rho_v$. Then the $i=k=0$ component of Einstein's equations
becomes
\begin{equation}
3 \left( \frac{\dot a}{a} \right)^2=\aleph (\rho_c+\rho_v)
\end{equation}
where $\aleph=8 \pi G$ and the equation for the scalar curvature
gives
\begin{equation}
6 \left[ \frac{\ddot a}{a}+ \left( \frac{\dot
a}{a}\right)^2\right]=\aleph (\rho_c+\rho_v-3 p_v)= \aleph
[\rho_c+\rho_v(1-3 w)].
\end{equation}
From these equations we derive
\begin{equation}
\label{decel} q=-\frac{\ddot a a}{\dot a^2}=\frac{1}{2}\left(
1+\frac{3\delta_v w}{1+\delta_v} \right);
 \quad \delta_v=\rho_v/\rho_c.
\label{q}\end{equation} The proportionality constant $w$ is
related to the deceleration parameter $q$ by
\begin{equation}
\label{w}
w=\left( \frac{2q-1}{3} \right) \left( \frac{\delta_v+1}{\delta_v} \right).
\end{equation}

The observed value of $q$ at red-shift parameter $z=a_0/a-1 \ll 1
$ defines the values of $y_0$ and $y_0^\prime$ in Eq.~(\ref{00})
in the present point $a_0=1$. The data \cite{Sahni} seem to imply
that $w<0$ corresponding to the inflation stage.

\section{Quintessence Model with $f=A \rho^n$}

Instead of our fluctuation-generated additional terms in
(\ref{addterm}), a model with an additional term $\Delta
L(R)=f(R)=A\rho^n\propto R^n$ has been considered in
Ref.~\cite{Capoz}. Let us review this case briefly to see the
difference with respect to our theory. The basic equation for dark
energy from (\ref{00}) reads
\begin{equation}
\label{00n}
A \left( \frac{\gamma}{a^2}\right) \left( \frac{y^\prime}{a^3}\right)^{n-2} \left[
n(n-1) y y^{\prime \prime}+\frac{1-n}{2}y^{\prime 2}+(4-3n)n\frac{y y^\prime}{a}\right]=\Omega_c a -y,
\end{equation}
where $\gamma=(-3)^{n-1}$, $y^{\prime}=dy/da$, where $A$ is still
arbitrary. Variation of the Lagrangian $L(R)=f(R)$ gives, instead
of Eq. (\ref{T}),
\begin{equation}
\label{T1}
\hat{T}_{i}^k=\aleph T_i^k,
\end{equation}
where $T_i^k$ is energy-momentum tensor of usual matter. The
authors in Ref.~\cite{Capoz} analyze these equation without taking
into account of usual matter. In this approximation, the solution
of our theory would be obtained by setting the tensor in
Eq.~(\ref{T}) equal to zero: $\hat{T}_{i}^k=0$. For the choice
$f(R)=A R^n$, the corresponding equation is obtained by setting
the brackets in Eq.~(\ref{00n}) equal to zero. The solution is
\begin{equation}
y=a^k; \quad k=\frac{(4n-5)2n}{(n-1)(2n-1)}. \end{equation} By
Eq.~(\ref{ya}), this is corresponds to the following expression
for the scale factor
\begin{equation}
\label{a_pow}
a=\theta^{\alpha};\quad \alpha=\frac{(n-1)(2n-1)}{2-n}.
\end{equation}
This solutions has a constant deceleration parameter $q$. Further,
to comply with the observable $H_0$, Ref.~\cite{Capoz} determined
certain intervals of $n$ with different ages of the universe. This
allowed him to find the deceleration parameter $q$ and parameter
$w$ (see Section \ref{basic}) as:
\begin{equation}
q=(1/\alpha-1);\quad w=(2-3\alpha)/3\alpha.
\end{equation}
Present data force us to reject such a pure power dependence of
the scale factor in (\ref{a_pow}). The latest data show
 that at present ($z=0$),
 the universe is accelerating its expansion
($w\approx -1$) but at $z=1$ there was no inflation ($w\approx
0$). Apart from that, the influence of cold matter cannot be
neglected~\cite{Star}.

Thus we need another ansatz for $f(\rho)$ to account for present
observations. From (\ref{00n}) we see that a simple power in the
asymptotic solutions is absent for $n=2$ and $n=4/3$. These happen
to be the same parameters which were of special importance
 in a previous
theory of the early universe. At $n=4/3$, one of us \cite{Gur2}
has obtained for the first time a
 cosmological model without
a singularity. That model passes through a regular minimum, had
inflationary stage and tends asymptotically to the classical
Friedmann solutions. In addition, the usefulness of an additional
term $R^2$ in the models of the early universe has been pointed
out before. Therefore we analyze
 the possibility of using of such
powers for construction of models with variable parameters $q$ and
$w$.

\section{Models with $n=2$ and $n=4/3$}
\label{R_2}
\subsection{$f(\rho)=-A\rho^2$}
This model was often used in the theory of the early universe.
 It describes the stage of fast
oscillations of $a(\theta)$ (the so called scalaron stage, which was introduced by A. Starobinsky in \cite{Star1}). The damping of such
oscillations was connected with creation of unstable particles and
filling of the early universe by a hot plasma.

Here we want to analyze this model in the opposite regime when the
period of oscillations is commensurable with age of the universe
$1/H_0$. One can easily see the oscillations of the model by
inserting the specified form of quintessence in (\ref{einst}) and
(\ref{T})
\begin{equation}
\frac{d^2\rho}{d\theta^2}+3\frac{\dot{a}}{a}\frac{d\rho}{d\theta}+(\rho+\Omega_c a)=0.
\end{equation}
The scalar curvature performs
 oscillations near the value of
$\rho$ which corresponds to the model of cold dust matter in the
Friedmann universe.

More conveniently, we may use the $i=k=0$ component of the
generalized gravitational equation (\ref{00n})
\begin{equation}
6 A \left[ y^{\prime \prime}y-y^{\prime 2}/4-2y y^{\prime}/a
\right]=-a^2(y-\Omega_c a).
\end{equation}
Introducing the quadratic scale variable $\xi= a^2$, this becomes
\begin{equation}
\label{geq}
24 A \left[ y \frac{d^2y}{d\xi^2}-\frac{1}{4}\left( \frac{dy}{d\xi} \right)^2-\frac{y}{2\xi}\frac{dy}{d\xi}
\right]=\Omega_c\sqrt{\xi}-y.
\end{equation}
 The initial conditions are
 specified at $a_0=1$, where the reduced Hubble parameter
$\dot{a_0}/a_0$ is unity. This condition defines $y_0=1$ at given
$\xi_0=1$ [see Eq.~(\ref{ya})].

At $y_0=1$, the value of derivative $(dy/da)_0$ is determined
 by observations in
 the following way. Using the observable
estimation for $\Omega_v=0.7$ and $\Omega_c=0.3$, the value of
$\delta_{v0}$ in Eq.~(\ref{q}) is approximately equal to 2. The
observations for $z<<1$ show that $w\approx -1$ and, using
(\ref{q}), we find
\begin{equation}
 q_0\approx\frac{1}{2}(1-2)= -1/2.
\end{equation}
At $\xi_0=1$,  we obtain from (\ref{decel})
\begin{equation}
\left( dy/d\xi \right)_0\approx 3/2.
\end{equation}
The only open parameter left is  $A$ in Eq.~(\ref{geq}). It
specifies the amount of dark energy in the universe.

The solution  $y(\xi)$  in the interval $0.01<\xi<1$ is presented
in Fig. 1. The age of the universe $T$ is defined by the value of
the Hubble parameter
\begin{equation}
\frac{\dot{a}}{a}=h(\xi)=\sqrt{y(\xi)}/\xi, \quad \theta=T H_0=
\int_{\epsilon}^{1} \frac{d \ln{\xi}}{2
h(\xi)}=\int_{\epsilon}^{1} \frac{d\xi}{2 \sqrt{ y(\xi)}}, \quad
\epsilon\rightarrow 0.
\end{equation}
Note that Eq.~(\ref{geq}) yields for  $\xi\rightarrow 0$ the
regular behavior $y\rightarrow$ const. Thus the integral at
$\epsilon \rightarrow 0$ has no singularity.

 The age $T$ depends on
choice of the parameter $A$ in (\ref{geq}). The satisfactory value
$T \approx 0.83/H_0\approx 11.4 $ Gyr is reached at $A\approx 1$.
For a further analysis of the model and comparison  with
observable data see Section \ref{disc}.
\begin{figure*}
\begin{center}
\begin{picture}(48.5,222.75)
\put(-210,6){\includegraphics[width=\textwidth]{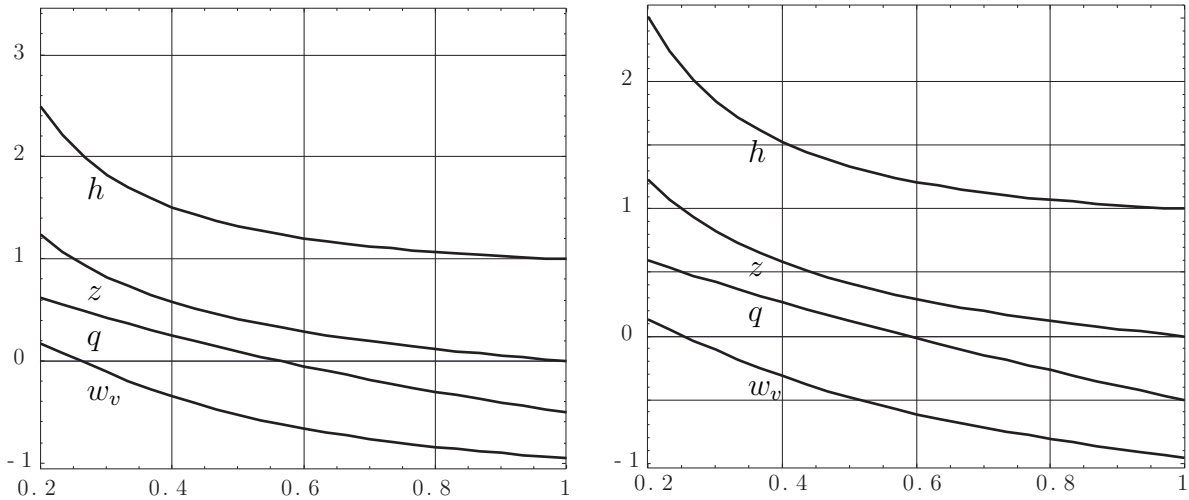}}
\put(70,45){$w_v$}
\put(70,73){$q$}
\put(70,90){$z$}
\put(70,133){$h$}
\put(-180,43){$w_v$}
\put(-180,64){$q$}
\put(-180,81){$z$}
\put(-180,119){$h$}
\end{picture}
  ~\\
 \caption{\small For the models with $n=2$ (left graph)
and with $n=4/3$ (right graph)
 the following parameters subject to $\xi$ are shown.
 }\label{fig1}
\end{center}
\end{figure*}

\subsection{$f(\rho)=-A\rho^{4/3}$}

The special feature of the model with $n=4/3$ consists in absence
of the scale factor $a$ in explicit form in the Einstein's
equations if matter is neglected \cite{Gur2}. This allows one to
find the general solution of Eq. (\ref{00n}) with given $n$. One
can show that de Sitter's solution arises in the limit $a\gg 1$.
The curvature of this limiting solution is determined
 by the parameter
$A$.

The evolution equation for this case is from (\ref{00n}):
\begin{equation}
\label{geq1} y \frac{d^2y}{d\xi^2}-\frac{3}{8}\left(
\frac{dy}{d\xi}\right)^2+\frac{1}{2}\frac{y}{\xi}\frac{dy}{d\xi} =
-\frac{9}{16 }\frac{1}{3^{1/3} A}  \left(\frac{2}{\xi} \frac{d
y}{d\xi}\right)^{2/3} \left[\Omega_c \sqrt{\xi}-y \right].
\end{equation}

By analogy with previous subsection, we obtain for this model (see
Fig. 1) an age of the universe $T \approx 0.79/H_0\approx 10.8 $
Gyr with a parameter $A=0.33$.

Leaving the analysis of the solution to Section \ref{disc}, let us
consider here only  the following issue. For $\xi
>1$, the influence of cold matter becomes negligible.
Then Eq.~(\ref{geq1}) has a first integral. It can be easily seen
from (\ref{00n}) for $n=4/3$:
\begin{equation}
y \frac{d^2 y}{d a^2}-\frac{3}{8}\left( \frac{dy}{d a}\right)^2=D
\left( \frac{dy}{d a}\right)^{2/3} y, \quad D=\frac{4}{9}
(3)^{-1/3}/A.
\end{equation}
The first integral is
\begin{equation}
\label{fi}
\left( \frac{dy}{d a}\right)^{4/3}=\frac{8}{3} D(y+C
\sqrt{y}),
\end{equation}
where $C$ denotes an integration constant. For $C>0$, this
equation has the asymptotical solution $y\approx \gamma a^4,
\gamma=2 (D/3)^3$, which coincides with the de Sitter solution
\begin{equation}
y=(a \dot a)^2=\gamma a^4; \quad a=a_0 exp{(\sqrt{\gamma} \theta)}.
\end{equation}

For  $C<0$, the solution possess a scale parameter $a$
 where $dy/da=0$.
Beyond that, $y$ starts decreasing. The solution reaches a point
$y\rightarrow 0$. By definition,
 $y$ can never be
be negative. Analytic continuation of the solution beyond the
specified point leads to an increase of $y$ at decreasing of $a$.
As  $a$ approaches zero we obtain a new singularity \cite{Breiz}.
The zero of $y$ corresponds to a maximum of the scale factor $a$
with subsequent contraction of the universe ending in a
singularity.

This is an example for a {\rm big crunch} (see \cite{Sahni} and
references therein). It is important to note that regime of a
crunch is realized here for a flat cosmological model.

Note that a contraction of cosmological model to a point cannot
really be described by a homogenous model. In the limiting regime,
the universe becomes certainly inhomogeneous and anisotropic
because of fast growth of perturbations. Using of the solution of
Eq. (\ref{fi}) with parameters $D$ and $C$ corresponding to observable data
for $z\ll 1$ guarantees an absence of a crunch in future ($C>0$).

\section{Discussion}
\label{disc}

After fitting the model's parameters to the present observable
data - acceleration of the universe, the Hubble parameter at
red-shift parameter $z=0$,
 and the age of the
universe - there are no free parameters in the model. Its
predictions for large $z$ can be compared with observations.

As can be seen in Fig. 1 (in case of $n=2$),  the acceleration of the expansion
 changes
 at $z\approx 0.46$ to a
deceleration ($w\approx -0.72, \, q\approx 0$). Near $z=1$ the
variant of dust-like dark energy ($w=0$) is realized. It
corresponds to latest observational data \cite{Sahni}.

A special feature of the model is that for $a\rightarrow 0$, the
variable  $y=(\dot a a)^2$   tends to a constant, which is
equivalent to an evolution of the universe filled by hot matter
($p=\epsilon/3$) in the Friedmann cosmology. When the solution is
oscillating, the inflation at $z\approx 0$ cannot be eternal
although the period of oscillations is comparable with the age of
universe.

In case of $n=4/3$, the crossover
from a
decelerated expansion
to an accelerated one takes place at $z\approx 0.30$ when
$w\approx -0.60$.

It is interesting to compare our model with the simplest LCDM model (for review see \cite{Sahni}).
Using
the notation
of Section (\ref{basic}), we  obtain
\begin{equation}
\label{lcdm_y}
y=\Omega_c \xi^{1/2}+\Omega_{v} \xi^2
\end{equation}
with the same values $\Omega_c\approx 0.3$ and $\Omega_v\approx 0.7$

Our results show the following: the age of the universe is near $1/H_0\approx 13.7$ Gyr.
This age is larger than the age from the pure $\rho^2$-model ($\approx 11.4$ Gyr) and from the $\rho^{4/3}$-model ($\approx 10.8$ Gyr).
The parameter $w$ of
 the LCDM model calculating with use of (\ref{w})
has the value  $-1$. We may use
Eq.~(\ref{q})
to determine the deceleration parameter
$q$  and compare it with the $\rho^2$ and $\rho^{4/3}$ models.
At $z=1 $, the deceleration parameter  is $q\approx 0.2$, whereas
 the $\rho^2$ and $\rho^{4/3}$ models have $q\approx 0.5$. This value is closer
to observable data. At $z\approx 0.5$, the parameter $q$ in the $\rho^2$-model is close to zero and in the LCDM model
$q\approx -0.1$. Apparently,
the $\rho^2$-model corresponds better to observations.
The LCDM model is mathematically very simple.
But it leaves the value
of the $\Lambda$-term an unexplained fundamental constant.
 For dynamical
models of the $\Lambda$-term (for example, the $\rho^2$-model) this value evolves from a
large Planck value
in the early universe to small value at present.

\section*{Acknowledgements}

We are grateful to A. Starobinsky for pointing of the paper
\cite{Capoz} and possibility of using of our previous papers
\cite{Breiz,Gur1,Gur2} in the given ansatz.
Work was partially supported by DAAD, by
the European Network
COSLAB, and by the ISTC project KR-677.

\end{document}